\documentclass[a4paper]{article}
\title{A mechanistic macroscopic physical entity with a three-dimensional
Hilbert space description\footnote{Published as: Aerts, D., Coecke, B., D'Hooghe, B. and Valckenborgh, F, 1997,
``A mechanistic macroscopic physical entity with a three-dimensional
Hilbert space description", {\it Helv. Phys. Acta}, {\bf 70}, 793.}}
\author {Diederik Aerts, Bob Coecke, Bart D'Hooghe and Frank Valckenborgh}

\date {}

\newcommand{\sep}{\hskip -2pt \makebox[11pt]{$\bigcirc$ \hskip -11.5 pt
\raise 0.1 pt \hbox{$\wedge$}}}

\newcommand{\smallsep}{\footnotesize\hskip -2pt \makebox[11pt]{$\bigcirc$
\hskip -10 pt \raise 0.1 pt \hbox{$\wedge$}}}

\usepackage{graphics,epsfig}
\usepackage{indentfirst}
\usepackage[psamsfonts]{amssymb}

\font\mediumroman=cmr10 at 8pt
\def\be{\begin{equation}}
\def\ee{\end{equation}}
\def\bea{\begin{eqnarray}}
\def\eea{\end{eqnarray}}

\newcommand{\real}{{\mathbb R}}

\begin{document}

\maketitle

\centerline{Center Leo Apostel (CLEA) and}
\centerline{Foundations of the Exact Sciences (FUND),}
\centerline{Brussels Free University,}
\centerline{Pleinlaan 2, B--1050, Brussels.}
\centerline{diraerts@vub.ac.be, bocoecke@vub.ac.be,}
\centerline{bdhooghe@easynet.be, fvalcken@vub.ac.be}

\begin{abstract}
\noindent It is sometimes stated that Gleason's theorem prevents the construction of
hidden-variable models for quantum entities described in a more than two-dimensional Hilbert space.
In this paper however we explicitly construct a classical (macroscopical) system that can be
represented in a three-dimensional real Hilbert space, the probability structure appearing as the
result of a lack of knowledge about the measurement context. We briefly discuss Gleason's theorem
from this point of view.
\end{abstract}

\section{Introduction} 
\noindent Even after more than 60 years there remain many
 problems on the 'understanding' of quantum mechanics. From the
 early days, a main concern of the majority of physicists reflecting on
 the foundations of the theory has been the question of understanding the nature of the quantum
 probability. At the other hand, it was a problem to understand the appearance of
 probabilities
 in
 classical theories, since we all agree that it finds its origin in a lack of knowledge
 about a deeper deterministic reality. The archetypic example is found in
 thermodynamics, where the probabilities
 associated with macroscopic observables such as pressure,
 volume, temperature, energy and entropy are due to the fact
 that the 'real' state of the entity is characterized
 deterministically by all the microscopic variables of positions and
 momenta of the constituting entities, the probabilities
 describing {\it our} lack of knowledge about the microscopic state of the
 entity. The variables of momenta and positions of the individual entities
 can be considered as 'hidden variables', present in the
 underlying reality. This example can stand for many of the attempts that
 have been undertaken to explain the notion of quantum probability, and the underlying theories are
 called 'hidden variable' theories.   In general, for
 a hidden variable theory, one aims at constructing a theory of an
 underlying deterministic reality, in such a way that the quantum
 observables appear as observables that do not reach this underlying
 'hidden' reality and the quantum probabilities finding their origin in a lack of knowledge about this
 underlying reality.  

 Von Neumann $^{1}$ gave a first impossibility
 proof for hidden variable theories for quantum mechanics.
 It was remarked by Bell $^{2}$ that in the proof of his
 No-Go theorem, Von Neumann had made an assumption that was not
 necessarily justified, and Bell explicitly constructs a hidden
 variable model for the spin of a spin-${1\over 2}$ quantum
 particle. Bell also criticizes the impossibility proof of Gleason
 $^{3}$, and he correctly points out the danger of demanding extra
 'mathematical' assumptions without an exact knowledge on their physical meaning. Very specific
attention was
 paid to this danger in the study of Kochen and Specker $^{4}$, and
 their impossibility proof is often considered as closing the
 debate. We can state that each of these impossibility proofs
 consists in showing that a hidden variable theory (under certain
 assumptions) gives rise to a certain mathematical structure for the
 set of observables of the physical system under consideration, while
 the set of observables of a quantum system does not have this
 mathematical structure. Therefore it is impossible to replace
 quantum mechanics by a hidden variable theory (satisfying the
 assumptions). To be more specific, if one works in the category of
 observables, then a hidden variable theory (under the given
 assumptions) gives rise to a commutative algebraic structure for
 the set of observables, while the set of observables of a quantum
 system is non-commutative. If one works in the category of
 properties (yes-no observables) then a hidden variable theory
 (satisfying the assumptions) has always a Boolean lattice structure
 for the set of properties while the lattice of properties of a
 quantum system is not Boolean. If one works in the category of
 probability models, then a hidden variable theory (satisfying the
 assumptions) has always a Kolmogorovian probability model for the
 set of properties while the quantum probability model is not
 Kolmogorovian. Most of the mathematically oriented physicists, once
 aware of these fundamental structural differences, gave up the hope
 that it would ever be possible to replace quantum mechanics by a
 hidden variable theory. However, it turned out that the state of affairs was even 
 more complicated than the structural differences in the different
 mathematical categories would make us
 believe.  We have already mentioned that the No-Go theorems for
 hidden variables, from Von Neumann to Kochen and Specker, depended
 on some assumptions about the nature of these hidden variable
 theories. We shall not go into details about the specific
 assumptions related to each specific No-Go theorem, because
 in the mean time it became clear that there is one central assumption
 that is at the core of each of these theorems: the hidden variables have to be hidden variables of
 the {\it state} of the physical entity under consideration and specify a deeper underlying
 reality of the physical entity itself, independent of the specific
 measurement that is performed. Therefore we shall call them state hidden variables. This
 assumption is of course inspired by the
 situation in thermodynamics, where statistical mechanics is the
 hidden variable theory, and indeed, the momenta and positions of
 the molecules of the thermodynamical entity specify a deeper
 underlying reality of this thermodynamical entity, independent of
 the macroscopic observable that is measured.  It was already remarked that
 there exists always the mathematical possibility to construct so-called
 contextual hidden variable models for
 quantum particles, where one allows the hidden variables to depend
 on the measurement under consideration (e.g. the spin model proposed
 by Bell $^{2}$). For the general case we refer to a theorem proved by Gudder $^{5}$.
 However, generally this kind of theories are
 only considered as a mathematical curiosum, but physically rather
 irrelevant. Indeed, it seems difficult to conceive from a physical
 point of view that the nature of the deeper underlying reality of
 the quantum entity would depend on the measurement to be performed.
 To
 conclude we can state that : (1) only state hidden variable theories
 were considered to be physically relevant for the
 solution of the hidden variable problem, (2) for non-contextual
 state hidden variable theories the No-Go theorem of Kochen and
 Specker concludes the situation; it is not possible to construct a
 hidden variable theory of the non-contextual state type that
 substitutes quantum mechanics.   

 What we want to point out is that,
 from a physical point of view, it is possible to
 imagine that not only the quantum system can have a deeper
 underlying reality, but also the physical measurement process for
 each particular measurement. If this is true, then the physical origin of
 the quantum probabilities could be connected with a lack of knowledge
 about a deeper underlying reality of the measurement process. 
 In $^{6,7,8}$ this idea was explored and it has been shown that such a lack of knowledge
 gives indeed rise to a quantum structure
 (quantum probability model, non-commutative set of observables, non-Boolean lattice of
 properties). This uncertainty about the
 interaction between the measurement device and the physical
 entity can be eliminated by introducing hidden variables that
 describe the fluctuations in the measurement context. However, they are not state
 hidden variables, they rather describe an
 underlying reality for each measurement process, and therefore they
 have been called 'hidden measurements', and the corresponding
 theories 'hidden measurement theories'. 
 Suppose that weperform a
 measurement $e$ on a physical system $S$ and that there is a lack of
 knowledge on the measuring process connected with
 $e$, in such a way that there exist 'hidden
 measurements' $e_\lambda$, where each $e_\lambda$ has the same
 outcome set as $e$, and each $e_\lambda$ is deterministic, which
 means that for a given state $p$ of the system $S$, for each
 $\lambda$ the hidden measurement $e_\lambda$ has a determined
 outcome. Now the fundamental idea is that each time when the
 measurement $e$ is performed, it is actually one of the $e_\lambda$, each with a certain
 probability, that
 takes place in the underlying hidden reality.
 In $^{6,7,8}$ it is shown that a hidden measurement model can be
 constructed for any arbitrary quantum mechanical system of finite
 dimension , and the possibility
 of constructing a hidden measurement model for an infinite dimensional quantum
 system can be found in $^8$. Although the models presented in these papers illustrate our point
 about the possibility of explaining the
 quantum probabilities in this way, there is always the possibility to
 construct more concrete macroscopic models, only
 dealing with real macroscopic entities and real interactions between
 the measurement device and the entities, that give rise
 to quantum mechanical structures. It is our point of view that
 these realistic macroscopic models are important from a physical
 and philosophical point of view, because one can visually perceive
 how the quantum-like probability arises. One of the authors 
 introduced
 such a real macroscopic model for the spin of a spin-${1\over 2}$
 quantum entity. When he presents this spin model for an
 audience,
 it was often raised that this kind of realistic
 macroscopic model can only be built for the case of a two-dimensional Hilbert space
 quantum entity, because of the theorem of Gleason and the paper of Kochen
 and Specker. Gleason's theorem is only valid for a Hilbert space
 with more than two dimensions and hence not for the two-dimensional
complex Hilbert space that is used in quantum mechanics
 to describe the spin of a spin-${1\over 2}$ quantum entity. In the
 paper of Kochen and Specker also a spin model for the spin of a
 spin-${1\over 2}$ quantum entity is constructed, and a real
 macroscopic realization of this spin model is proposed. They point out on different occasions that
 such a real model
 can only be constructed for a quantum entity with a Hilbert space
 of dimension not larger than two. The aim of this paper is to
 clarify this dimensional problem. Therefore we shall construct a
 real macroscopic physical entity and measurements on this entity
 that give rise to a quantum mechanical model for the case of a
 three-dimensional real Hilbert space, a situation where Gleason's
 theorem is already fully applicable. 
 We remark that one of the authors $^{10}$ presented a model
 for a spin-$1$ quantum entity that allows in a rather straightforward way a
 hidden measurement representation. Nevertheless, since he only considered a set of coherent spin-$1$
 states (i.e., a set of states that spans a three-dimensional Hilbert space, but that does not fill
 it) his model cannot be considered as a satisfactory counter argument
against the No-Go theorems.

In the first two sections, we briefly give the two-dimensional 
 examples of Aerts and Kochen-Specker and analyze their differences.
 In  section 4 we investigate the dimensional problem related to the
 possible hidden variable models. Afterwards, we  construct  a hidden
measurement model with a  mathematical
 structure for its set of states and observables that  can be
 represented in a three-dimensional real Hilbert space.

\section{The two-dimensional model}   
\noindent   
 The  physical entity that we consider is a point particle $P$
 that  can move on the surface of the unit sphere $S^2$.  Every unit vector ${\bf v}$ 
represents a state $p_v$ of
 the entity.  For every point ${\bf u}$ of $S^2$ we define a
 measurement $e_u$ as  follows : a rubber string between
 ${\bf u}$ and its antipodal point ${\bf -u}$ catches the particle $P$ that
 falls  orthogonally and sticks to it. Next, the 
 string breaks somewhere with a uniform probability density and the particle $P$ moves to one  of
 the points ${\bf u}$ or ${\bf -u}$, depending on the piece of 
 elastic it was attached to. If it arrives in ${\bf u}$ we will give the 
 outcome $o^u_1$ to the experiment, in the other case we will say 
 that the outcome $o^u_2$ has occurred. After the 
 measurement the entity will be in a new state: $p_u$  in the case of
 outcome $o^u_1$ and $p_{-u}$ in the other case.  Taking into account that the elastic
 breaks uniformly, it is easy to calculate the probabilities for the two results:  
$$P(o^u_1|p_v)= {1 + \cos \theta
 \over  2} = \cos^2 { \theta \over 2} $$    
$$P(o^u_2|p_v)={1 - \cos \theta
 \over 2 } = sin^2 { \theta \over 2 } $$
\noindent with $\theta = {\bf u} \cdot {\bf v}$.  
We have the same
 results for the probabilities associated with the spin
 measurement of a  quantum entity of spin-${1 \over 2}$ (see
 $^{2,8}$), so we can  describe our macroscopic
 example by the ordinary  quantum formalism where the set of states
 is given by the points  of a two-dimensional complex Hilbert space. 
  Clearly, we can also interpret this macroscopic example as a hidden variable model
 of the spin measurement of a quantum entity of spin-${1 \over 2}$.
 Indeed,
 if the point $\lambda$ where the string disintegrates is known, the
 measurement outcome is certain. The probabilities in
 this model appear because of our lack of knowledge of the precise
 interaction  between the entity and the measurement device. Every spin
 measurement  $e_u$ can be considered as a class of classical spin measurements 
$e_u^\lambda$ with determined outcomes, and the
 probabilities are the result of an averaging process.  In this example it
 is clear that the hidden  variable $\lambda$ is
 neither a variable of the entity under  study nor a variable
 pertaining to the measurement apparatus. Rather, it is a variable belonging
 to the measurement process as a whole.    
 
\section{The Kochen-Specker example}  
\noindent   In Kochen and Specker's model, again a point $P$ on a sphere represents the quantum
 state of the spin-${1 \over 2}$ entity. However, at the same time the
 entity is in a hidden state  which is represented by another point
 $T_P$ of $S_P^+$, the upper  half sphere with $P$ as its north pole,
 determined  in the following way.
 A disk $D$ of
 the same radius as the sphere  is placed perpendicular to the line
 $OP$ which connects $P$ with the  center $O$ of the sphere and
 centred directly above $P$. A particle  is placed on the disk that
 is now shaken ``randomly'', i.e., in  such a way that the
 probability that the particle will end up  in a region $U$ of the
 disk is proportional to the area of $U$. The  point $T$ is then the
 orthogonal projection of the particle. The  probability density
 function $\mu (T)$ is   $$ \mu (T) = \left\{\matrix{  {1 \over \pi}
 \cos \theta & 0 \leq \theta \leq {\pi \over 2} \cr  0 &  {\pi \over
 2} \leq \theta \leq \pi \cr  }\right. $$  where $\theta$ is the
 angle between $T$ and $P$.  If a measurement is made in the
 direction $OQ$ the outcome  ``spin  up'' will be found in the case
 that $T \in S_Q^+$ and ``spin  down'' otherwise. As a result of the
 measurement the new state  of the entity will be $Q$  in case of
 spin up and $-Q$ otherwise. The  new hidden state $T_Q$ is now
 determined as before, the
 disk being placed now at $Q$ if the new state is
 $Q$ and at $-Q$ if otherwise.  It can be shown that the same
 probabilities as for the quantum  spin-${1 \over 2}$ entity occur. 
  \noindent It is important to remark that the hidden
 variable here  pertains to the entity under study, as was made
 clear by using  the expression ``hidden state''. But is this really
 the case?  As we look closer we see that for every consecutive 
 spin measurement to reveal the correct probabilities, we need 
 each time a randomisation of the hidden state $T$. Thus every time 
 a measurement occurs the hidden variable has to be reset again.  In
 practice this means that for every measurement a new value of  the
 variable will be needed. Thus we can make the philosophical 
 important step to remove this ``hidden state'' from the entity  and
 absorb it within the context of the measurement itself,
 indeed a reasonable thing to do. Once this is done,  the
 analogy with the model of section 3 is obvious. But it is also
 clear that a new idea has been introduced, namely the shift  of the
 hidden variable from the entity towards the measurement  process.
 This is not only a new feature for a hidden variable  theory, but
 also a natural way out of the traps of the No-Go  theorems.    
 
\section{The Dimensional problems}  
\noindent   As was pointed out by several authors (see 
 $^{3,4,5,6,13,14}$), it is possible to prove that ``reasonable'' hidden variable theories don't exist
 for Hilbert spaces with a dimension greater
 than two. 
 Moreover, other arguments show the necessity for a proof of existence of a
 hidden variable model with a more than two-dimensional state space. There is
 for instance the theorem of Gleason which states that for  a propositional system corresponding
 to a three-dimensional real  Hilbert space there exists a unique
 probability function over the  set of propositions which satisfies
 some very plausible  properties. This means that every hidden
 variable theory (satisfying these assumptions) can only  reveal the same
 probabilities as the
 quantum probability function  and this would render the hidden
 variable theory redundant,  because no extra information can be
 gained. To prove that the No-Go theorems
 are too restrictive it is  thus necessary (but also sufficient!) to
 give one ``reasonable'' example with a three-dimensional Hilbert state space and
 this is exactly what we will do now.    

\section{The 3-Dimensional model}   
 \noindent   In this section we introduce a mechanistic 
 macroscopic physical entity with  a  three-dimensional Hilbert
 space quantum description.  Probably  there exist models that are
 much more elegant than the one we propose, because the explicit realization would be rather non-trivial, but for our
purpose it is sufficient to prove
that there exists at least one. 
Once again we remark 
 that the system that we present is not a
 representation of a quantum mechanical entity, but a macroscopic physical entity that
 gives rise to the same probability structure as one encounters in quantum mechanics.  
 First we propose the model and, for reasons of readability, we 
 present a geometrical equivalent in $\real^3$.  In this way we can easily prove the
equivalence between 
 the model and the quantum mechanical case. In section 5.3 we shall study
 the probability structure of the model.     
 
\subsection{The practical realization} 
   
 \noindent   The entity $S$ that we consider is 
 a rod of length 2 which is fixed in its center point $c$, both 
 sides of which have to be identified. The set of states of the entity, i.e.  the set of rays
 in Euclidean $3$-space, possibly characterized by  one of the two
 end points of the rod (denoted by $x_p$), will be  denoted by
 $\Sigma_S$. The measurement apparatus consists of three 
 mutual orthogonal rods, parallel with rays  $\widehat x_e^1,
 \widehat x_e^2, \widehat x_e^3$, fixed in 3-space. The
 entity and the measurement device are coupled for a measurement
 in the following way :  (see Fig. 1) :  
 \par\noindent $\bullet$
 Connection in $x_e^i$: the rod floats in  a slider which is fixed
 orthogonal to the rod of the measurement  apparatus.  
\par\noindent $\bullet$ Connection in $x_p$: the three  interaction-rods are fixed to one slider,
which
 floats on the   ``entity-rod''.     
\par\noindent $\bullet$ We also
 fix three rubber strings between the  entity-rod and the three rods of
 the measurement apparatus.  
\par\noindent The last ingredient that
 takes part in the  interaction is something we call a ``random
 gun''.  This is a gun, fixed on a slider that floats on and turns
 around the  entity-rod in such a way that:  
\par\noindent
 $\bullet$ The gun is shooting in a direction  orthogonal to the
 entity-rod.   \par\noindent $\bullet$ The movement and the
 frequency of  shooting are at random but such that the probability
 of   shooting a bullet in a certain direction, and from a certain 
 point of the entity-rod is uniformly distributed, i.e., the gun 
 distributes the bullets uniformly in all directions and from all  
 the points of the rod. If
 a bullet hits one of the connections, both the rod and string break, 
 such that the entity can start moving
 (there is one new degree of freedom), and it is clear that  the two
 non broken strings will tear the entity into the plane  of the
 measurement-rods to which it is still connected.

\vskip 0.3 cm
\includegraphics{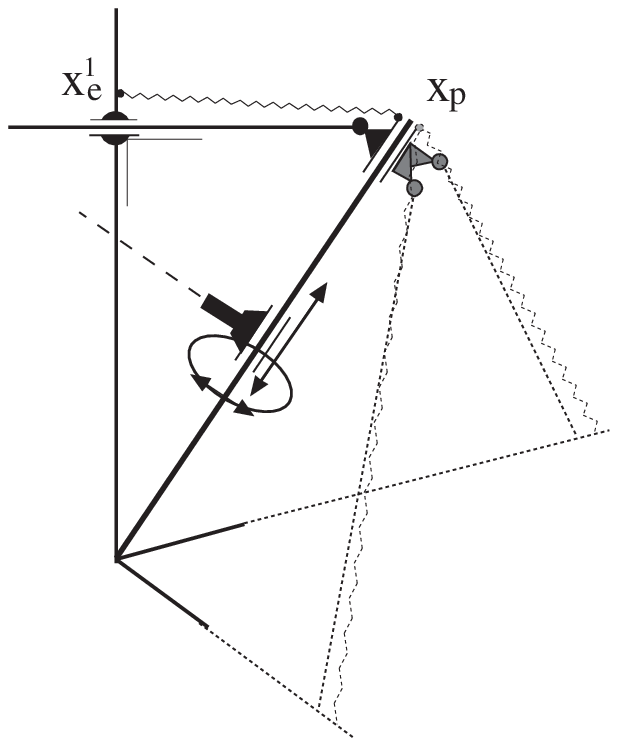}

\begin{quotation}
\noindent \mediumroman \baselineskip 8pt Fig. 1 :  Practical realization of the model.  With
 rods, sliders,  strings and a ``random gun" we construct a device with a mathematical
 structure equivalent to the one for a quantum entity with a three-dimensional real
Hilbert state space.
\end{quotation}

\subsection{A geometrical equivalent of the model} 
 
 \noindent To facilitate the calculation of the probabilities we 
 will describe what happens during the measurement from a 
 geometrical point of view.       
 We know that a state $p$ of the entity is characterized
 by the  angles $\theta_1, \theta_2, \theta_3$ between the rod and
 an  arbitrary selected set of three orthonormal axis in Euclidean 
 $3$-space $E^3$.  It is clear that this set of states 
 corresponds in a one-to-one way with the states of an entity 
 described in a three-dimensional real Hilbert space.    
 
\noindent The set of measurements to be performed on this  entity
 $S$ is characterized as follows.    Let $\widehat x_e^1, \widehat
 x_e^2, \widehat x_e^3$ be the three mutual orthogonal rays
 coinciding  with the rods of the
 measurement  apparatus. As a consequence, for a given state $p$,
 and a given  experiment $e$, we have the three angles  $\theta_1,
 \theta_2, \theta_3$ as representative parameters to characterize
 the state,  relative to the measurement apparatus. We denote by  $x_e^1, x_e^2,
 x_e^3$, the orthogonal projections of $x_p$ on the  three rays
 $\widehat x_e^1, \widehat x_e^2, \widehat x_e^3$, forming a set
 of points representative for the couple $(p , e)$.     
 The geometrical description of the measurement process goes as follows:    
 
 \noindent i) Every point $x_e^i$ is connected with $x_p$ by a 
 segment denoted by $[x_e^i,x_p]$ with length $sin\theta_i$. 
 Therefore the length of the projection
 of $[x_e^i,x_p]$ on the rod is $ cos({\Pi \over 2}
 -\theta_i).sin\theta_i=sin^2\theta_i$.         

 \noindent ii) Next, one of the connections $[x_e^i,x_p]$ breaks 
 with a probability proportional to the length of the projection of 
 $[x_e^i,x_p]$ on the rod (In Fig. 4 and Fig. 5 we suppose that 
 $[x_e^1,x_p]$ breaks).  The rod rotates into the plane  of the two remaining
 points $x_e^j,x_e^k$, to which it is still connected,  and such
 that the point $x'_p$, the projection of $x_p$ on  the
 $\widehat{x_e^j x_e^k}$-plane, lies on the rod.  As a  consequence,
 the connections $[x_e^j,x'_p]$ and $[x_e^k,x'_p]$  are still
 orthogonal to the corresponding axes $\widehat x_e^j$  and
 $\widehat x_e^k$.      

\vskip 0.4 cm
\includegraphics{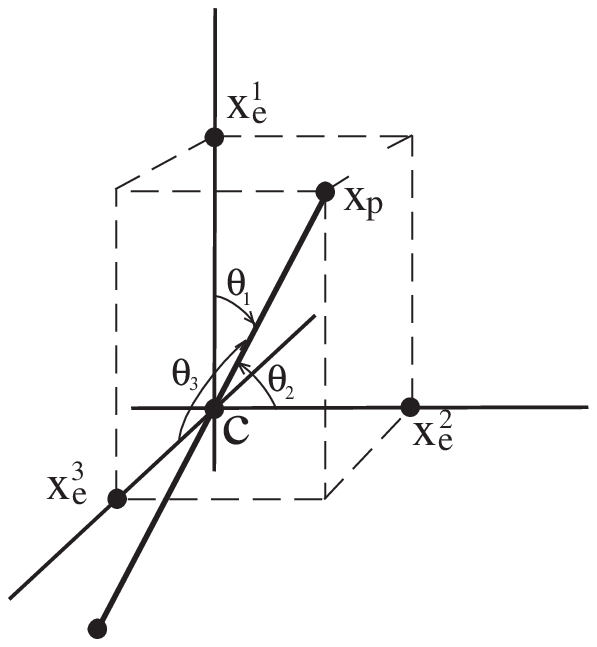}

\begin{quotation}
\noindent \mediumroman \baselineskip 8pt Fig. 2 : The  states of the classical mechanistic  
  entity, a rod in  Euclidean $\scriptstyle 3$-space, represented
 by $\scriptstyle x_p$, one of the two end points of the rod.  Thus,
 the different  states $\scriptstyle p$ of the entity are
 represented by the   angels $\scriptstyle \theta_1, \theta_2,
 \theta_3$ between the  rod and three mutual orthogonal rays 
 $\scriptstyle \widehat x^1_e, \widehat x^2_1, \widehat x^3_1$ , 
 representative for a measurement
 $\scriptstyle e$.  $\scriptstyle x^1_e, x^2_e, x^3_e$, the
 orthogonal projections of  $\scriptstyle x_p$ on the three rays 
 are
 thus representative  for the couple $\scriptstyle (p , e)$.
\end{quotation}
 
\vskip 0.2 cm
\includegraphics{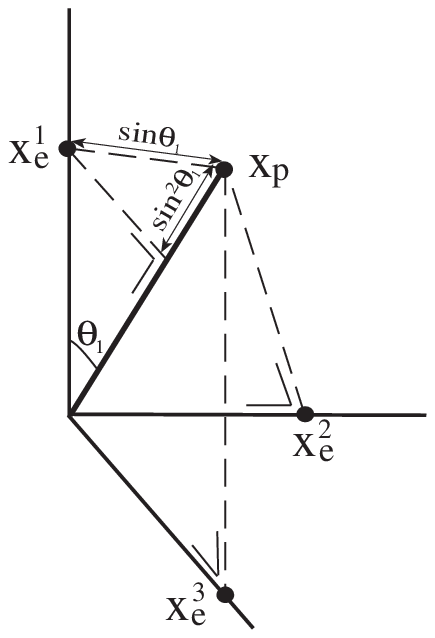}

\begin{quotation}
\noindent \mediumroman \baselineskip 8pt Fig. 3 : The  first step of the  measurement.  Every
 point $\scriptstyle x^i_e$ is connected with  $\scriptstyle x_p$ by
 a segment denoted by  $\scriptstyle [x^i_e,x_p]$.  The length of
 the projection of  $\scriptstyle [x^1_e,x_p]$ on the rod is
 $\scriptstyle sin^2\theta_1$.
\end{quotation}

\noindent iii) We proceed with this new, two-dimensional 
 situation characterized by the elements $\{x'_p,x_e^j,x_e^k\}$ as before,
 denoting the angle between $x'_p$ 
 and $x_e^j$ as $\theta'_j$.  One of the segments, 
 $[x_e^j,x'_p]$  or $[x_e^k,x'_p]$, seizes to exist, again with a probability 
 proportional to the length of the projection of this segment on 
 the rod, equal to $sin\theta_i.sin^2\theta'_j$. Ultimately, the rod rotates
 towards and stabilises at the third ray, to
 which it is still connected.      

\noindent The global process can thus  be seen as a
 measurement $e$, with three possible outcomes  $o_e^1,o_e^2,o_e^3$,
 on an entity $S$ in a state $p$.    

\includegraphics{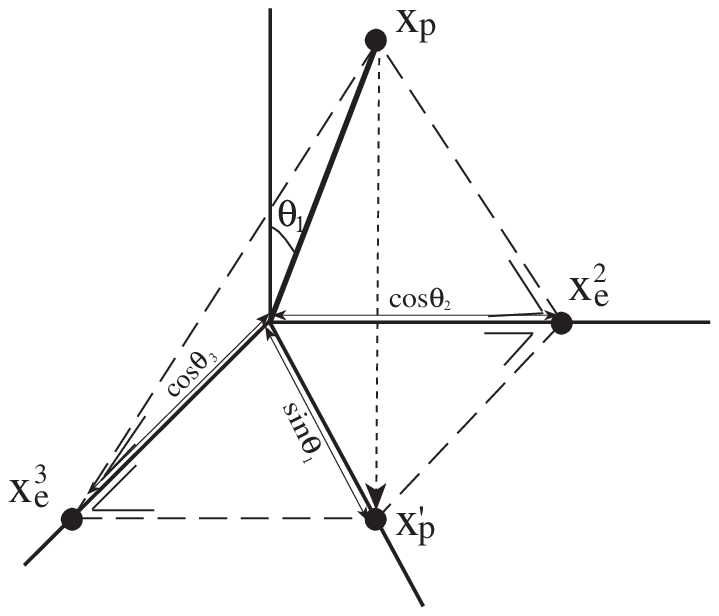}

\begin{quotation}
\noindent \mediumroman \baselineskip 8pt Fig. 4 : The  second step of the measurement.  One of
 the connections,  $\scriptstyle [x^1_e,x_p]$, breaks with
 probability proportional  to the length of the projection of
 $\scriptstyle [x^1_e,x_p]$ on  the rod.  The rod rotates into the
 plane of the two points  $\scriptstyle x^2_e,x^3_e$ in such a way
 that the connections  $ \scriptstyle [x^2_e,x'_p]$ and 
 $\scriptstyle [x^3_e,x'_p]$ are still orthogonal to the
 corresponding axes  $\scriptstyle \widehat x^2_e$ and 
 $\scriptstyle \widehat x^3_e$.
\end{quotation}

\vskip -2.5 cm

 \includegraphics{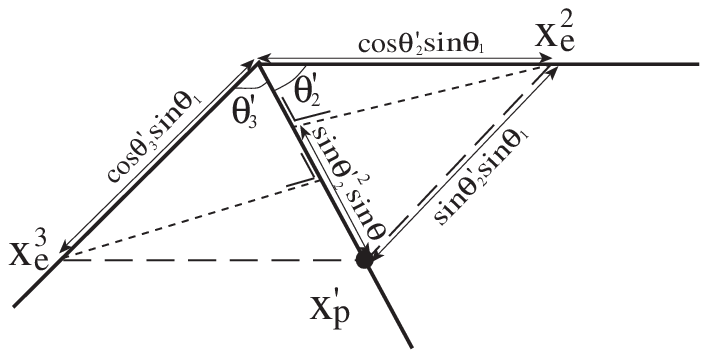}

\begin{quotation}
\noindent \mediumroman \baselineskip 8pt Fig. 5 : We  proceed with $\scriptstyle \{
 x'_p,x^2_e,x^3_e\}$ as we did with  $\scriptstyle \{
 x_p,x^1_e,x^2_e,x^3_e\}$.  One of the two  existing connections
 breaks with probability proportional to the  length of the
 projection of the corresponding segment on the rod, equal to $\scriptstyle
sin\theta_1.sin^2\theta'_2$.
\end{quotation}

\subsection{The probability structure of the model} 
 
 \noindent   After this geometrical representation of our 
 model it becomes very easy to calculate the probability
 to obtain  an outcome $o_e^k$, equivalent with neither obtaining
 $o_e^i$ nor
 $o_e^j$, and thus with the breaking of these two connections. 
 Suppose that first $[x_e^i,x_p]$ breaks and then
 $[x_e^j,x'_p]$. Since $cos^2\theta_i+cos^2\theta_j+cos^2\theta_k=1$
 we have 
 $sin^2\theta_i+sin^2\theta_j+sin^2\theta_k=2$. So we find $(sin^2\theta_i)/2$ for 
 the probability for the breaking of $[x_e^i,x_p]$.    Since
 $cos^2\theta'_j+cos^2\theta'_k=1$ we have 
 $sin^2\theta'_j+sin^2\theta'_k= 
 1$.  Thus $sin^2\theta'_j$ is the conditional probability
 for  the breaking of $[x_e^j,x'_p]$ supposing that the connection 
 between $x_e^i$ and the rod broke first.  This yields
 ${{1\over 2}sin^2\theta_i.sin^2\theta'_j  = {1\over
 2}sin^2\theta_i.cos^2\theta'_k    = {1\over
 2}sin^2\theta_i.({cos\theta_k\over sin\theta_i})^2    = {1\over
 2}cos^2\theta_k }$ for the requested probability. Analogously, we find the same result for the
 probability  that  first $[x_e^j,x_p]$ and then
 $[x_e^i,x'_p]$ breaks.   
 
\noindent Therefore we find: 
 $$P(o_e^k|p)=\cos^2\theta_k$$  
 
 \noindent where $\theta_k$ is the angle between $x_p$ and $x_e^k$, the eigenstate 
 with eigen-outcome $o_e^k$ of the measurement $e$ on the entity 
 $S$.

 \noindent Now we are able to compare the probability structure
 associated with our model with the one
 encountered in quantum mechanics.  
 For a three-dimensional real Hilbert space ${\cal H}_{{\real}^3}$
 we can write a self-adjoint operator $H_e$  with $\{\widehat
 x_e^1, \widehat x_e^2, \widehat x_e^3\}$  a set of mutual
 orthogonal eigen-rays and $\{o_e^1, o_e^2, o_e^3\}$ the
 corresponding eigenvalues (some of them may be equal), as $H_e =
 \sum_{i=1}^{i=3}o_e^i\ E_{\widehat x_e^i}$,  where $E_{\widehat
 x_e^i}$ is the projector on the ray $\widehat  x_e^i$. Therefore, we have for
 every ${o_e^i}$, eigen-outcome of a  measurement $e$ and associated
 with an eigenstate represented by  a ray $\widehat x_e^i$, and for
 every state $p$ of the system,  represented by a ray $\widehat
 x_p$:  $$P(o_e^i|p)= \vert <x_e^i \vert\  x_p>
 \vert^2 = \cos^2\theta_i$$    \noindent where $\theta_i$ is the
 angle between the rays   $\widehat x_e^i$ and $\widehat x_p$.  It is therefore clear that the
entity in our model
 corresponds in a one-to-one way with a quantum entity  described in
 a three-dimensional real Hilbert space.    
 
\section{Discussion}     
 
 \noindent In this paper we have presented a macroscopic device with a quantum-like probability
structure and state space. Since one can interpret this model as a hidden variable description for
a quantum entity, we can analyse the relationship with Gleason's theorem, which implies the
existence of a unique probability measure for a physical entity if its state space is a more than
two-dimensional separable Hilbert space ${\cal H}$ and if this probability measure satisfies some
reasonably looking a priori assumptions. For pure states Gleason's theorem takes the following
form: if $p:{\cal L(H)} \rightarrow [0,1]$ is a (generalised) probability measure, there exists a unit
vector $\psi \in {\cal H}$ such that $\forall\ P \in {\cal L(H)}: p(P) =\ <\psi\ \vert\ P \psi>$, 
with ${\cal L(H)}$ the lattice of closed subspaces of the Hilbert space. In
our case it asserts that the probability to obtain say $o_e^i$ necessarily takes the form that
was given above in this paper. Therefore it is implicit in the assumptions of the theorem
that the probabilities only depend on the initial and final state of the entity. However,
referring to our model we see that it is easy to invent other probability measures that actually
do depend on the intermediate states of the entity and therefore do not satisfy the
assumptions of Gleason's theorem. For instance, one can imagine that the random gun is absent and the
interaction rods break with a uniform probability density, resulting in the first probability being
proportional to $sin \theta_i$ in stead of $sin^2 \theta_i$. Since the hidden measurement approach is
obviously a contextual theory that keeps the Hilbert space framework for its state space, but
situates the origin of the quantum probability in the measurement environment, there is no need for
the existence of dispersion-free probability measures on ${\cal L(H)}$ as in the conventional
non-contextual state hidden variable theories.        
 
\section{References}  
 \medskip  
 \noindent  $^1$ J. Von Neumann, {\it Grudlehren}, Math. Wiss. 
 XXXVIII, 1932.  
 \smallskip 
 \noindent $^2$ J.S. Bell, Rev. Mod. Phys. {\bf 38}, 447, 1966. 
 \smallskip  

 \noindent $^3 $ A.M. Gleason, J. Math. Mech. {\bf 6}, 885, 1957. 
 \smallskip  
 \noindent $^4$ S. Kochen and E.P. Specker, J. Math. Mech. {\bf 
 17}, 59, 1967. 
 \smallskip
 \noindent $^5$ S.P. Gudder, J. Math. Phys {\bf 11}, 431 (1970). 
 \smallskip  
 \noindent $^6$ D. Aerts, {\it A possible explanation for the
 probabilities of quantum mechanics and a macroscopic situation that
 violates Bell inequalities}, in {\it Recent Developments in Quantum
 Logic}, eds. P. Mittelstaedt et al., in Grundlagen der Exacten
 Naturwissenschaften, vol. {\bf 6}, Wissenschaftverlag,
 Bibliographischen Institut, Mannheim, 235, 1984.  
 \smallskip  
 \noindent $^7$ D. Aerts, {\it A possible explanation for the
 probabilities of quantum mechanics}, J. Math. Phys. {\bf 27}, 202,
 1986.  
 \smallskip 
 \noindent $^8$ D. Aerts, {\it The origin of the non-classical
 character of the quantum probability model}, in {Information,
 Complexity and Control in Quantum Physics}, A. Blanquiere, et al.,
 eds., Springer-Verlag, 1987.    
 \smallskip 
 \noindent $^9$ B. Coecke, Found. Phys. Lett. {\bf 8}, 437 (1995).
\smallskip 
 \noindent $^{10}$ B. Coecke, Helv. Phys. Acta. {\bf 68}, 396 (1995).

\smallskip  
\noindent {$^{11}$}D. Aerts,
 Found. Phys. {\bf 24}, 1227 (1994).  
 \smallskip  
 \noindent {$^{12}$}D. Aerts, Int. J. Theor. Phys. {\bf 34}, 1165 
 (1995).  
 \smallskip  
 \noindent {$^{13}$}D. Aerts, {\it The Entity and Modern Physics} in 
 {\it  Identity and  Individuality of Physical Objects}, ed. T.
 Peruzzi, Princeton  University Press, Princeton,  (1995). 
 \smallskip 
 \noindent {$^{14}$}J.M. Jauch and C. Piron, Helv.  Phys.  Acta.
 {\bf 36},  827 (1963).     
 \smallskip 
 \noindent {$^{15}$}J.S. Bell, Rev. Mod.  Phys. {\bf 38}, 447 
 (1966).

 \end{document}